\documentclass[12pt,a4paper]{article}
\usepackage{epsfig, floatflt, amssymb, citesort}
\usepackage [small] {caption}

\begin{document}
\newcommand{\beqn}{\begin{equation}}
\newcommand{\eeqn}{\end{equation}}
\newcommand{\etag}{\eta\gamma}
\newcommand{\twog}{2\gamma}
\newcommand{\ee}{e^+e^-}
\newcommand{\pipi}{\pi^+\pi^-}
\newcommand{\pipipi}{\pi^+\pi^-\pi^0}
\newcommand{\pig}{\pi^0\gamma}
\newcommand{\etapipi}{\eta\pi^+\pi^-}
\newcommand{\eeetapipi}{\ee \to \etapipi}
\newcommand{\eeetag}{\ee \to \etag}
\newcommand{\rhop}{\rho\prime}

\date{}

\title{\Large
\bf \boldmath 
Study of the Process $\eeetag$ in c.m. Energy Range
600-1380~MeV at CMD-2}

\author{
R.R.~Akhmetshin\footnote{Budker
Institute of Nuclear Physics, Novosibirsk, 630090, Russia},
E.V.~Anashkin\footnotemark[1],
V.M.~Aulchenko\footnotemark[1]
\footnote{Novosibirsk State University, Novosibirsk, 630090, Russia},\and
V.Sh.~Banzarov\footnotemark[1],
A.~Baratt\footnote{University of Pittsburgh, Pittsburgh, PA 15260, USA},
L.M.~Barkov\footnotemark[1] \footnotemark[2], 
S.E.~Baru\footnotemark[1] \footnotemark[2], \and
N.S.~Bashtovoy\footnotemark[1],
A.E.~Bondar\footnotemark[1] \footnotemark[2],
D.V.~Bondarev\footnotemark[1], 
A.V.~Bragin\footnotemark[1],  \and
D.V.~Chernyak\footnotemark[1],
S.I.~Eidelman\footnotemark[1] \footnotemark[2], 
G.V.~Fedotovitch\footnotemark[1] \footnotemark[2],  \and 
N.I.~Gabyshev\footnotemark[1],
A.A.~Grebeniuk\footnotemark[1], 
D.N.~Grigoriev\footnotemark[1],
V.W.~Hughes\footnote{Yale University, New Haven, CT 06511, USA},\and
F.V.~Ignatov \footnotemark[1] \footnotemark[2], 
S.V.~Karpov\footnotemark[1],
V.F.~Kazanin\footnotemark[1],
B.I.~Khazin\footnotemark[1] \footnotemark[2],\and
I.A.~Koop\footnotemark[1],  
P.P.~Krokovny\footnotemark[1] \footnotemark[2], 
L.M.~Kurdadze\footnotemark[1],
A.S.~Kuzmin\footnotemark[1] \footnotemark[2],   \and 
I.B.~Logashenko\footnotemark[1], 
P.A.~Lukin\footnotemark[1],
A.P.~Lysenko\footnotemark[1],
K.Yu.~Mikhailov\footnotemark[1] , \and 
A.I.~Milstein\footnotemark[1] \footnotemark[2], 
I.N.~Nesterenko\footnotemark[1],
V.S.~Okhapkin\footnotemark[1],
A.V.~Otboev\footnotemark[1], \and
E.A.~Perevedentsev\footnotemark[1] \footnotemark[2], 
A.S.~Popov\footnotemark[1] \footnotemark[2],  
N.I.~Root\footnotemark[1] \footnotemark[2],
A.A.~Ruban\footnotemark[1], \and
N.M.~Ryskulov\footnotemark[1],
A.G.~Shamov\footnotemark[1],  
Yu.M.~Shatunov\footnotemark[1],
B.A.~Shwartz\footnotemark[1] \footnotemark[2], \and
A.L.~Sibidanov\footnotemark[1] \footnotemark[2],
V.A.~Sidorov\footnotemark[1], 
A.N.~Skrinsky\footnotemark[1],
V.P.~Smakhtin\footnotemark[1], \and
I.G.~Snopkov\footnotemark[1], 
E.P.~Solodov\footnotemark[1] \footnotemark[2],
P.Yu.~Stepanov\footnotemark[1],
A.I.~Sukhanov\footnotemark[1], \and
J.A.~Thompson\footnotemark[3], 
A.A.~Valishev\footnotemark[1], 
Yu.V.~Yudin\footnotemark[1],
S.G.~Zverev\footnotemark[1]
}

\maketitle

\begin{abstract}
The cross section of the process $\eeetag$ has been measured in the 
600-1380~MeV c.m. energy range with the CMD-2 detector.
The following branching ratios have been determined:
$$
 B(\rho \to \etag) =  (3.28\pm 0.37\pm 0.23) \cdot 10^{-4}\, ,
$$
$$
 B(\omega \to \etag) = (5.10 \pm 0.72\pm 0.34) \cdot 10^{-4}\, ,
$$
$$
 B(\phi \to \etag) =  (1.287\pm 0.013\pm 0.063) \cdot 10^{-2}\, .
$$
Evidence for the $\rho(1450) \to \etag$ decay has been obtained for the 
first time. 
\end{abstract}

\section{Introduction}
Radiative decays of the $\rho$, $\omega$ and $\phi$
and particularly their magnetic dipole transitions to the
$\etag$ final state have traditionally been a good laboratory
for various tests of theoretical concepts, from the quark
model and SU(3) symmetry to Vector Dominance Model (VDM) and
anomalous contributions~\cite{theor1,theor2,theor3,theor4,ben}. 
Among these decays only that of the $\phi$ meson has been well studied 
experimentally~\cite{pdg}. However, most of the previous measurements 
of the $\phi \to \etag$ decay were performed in a narrow energy 
range resulting in a large model uncertainty related to the description of 
the cross section outside the $\phi$ meson. This error can be decreased 
by measuring the cross section in a broader energy range.
Experimental information on the $\rho$ and $\omega$ 
decays is rather scarce~\cite{pdg}.    

The data samples of the $\etag$ 
production accumulated previously in the off-resonance region are 
negligible~\cite{nd} making impossible the determination of the 
cross section of the process $\eeetag$ which although small 
could contribute to the total hadronic  cross section and thus 
influence the precision of the calculation of the hadronic contribution
to (g-2)$_{\mu}$~\cite{ej}.

Large integrated luminosity collected in recent experiments at the VEPP-2M
$\ee$ collider in Novosibirsk allows qualitatively new analysis of the 
$\etag$ final state produced in $\ee$ annihilation.
CMD-2 has already published results of the $\phi \to \etag$ 
study in the $\eta \to \pi^+\pi^-\pi^0$ decay mode~\cite{egcmd}
whereas SND presented their determination of the  branching ratio of
the $\phi \to \etag$ decay via various decays modes of the 
$\eta$~\cite{egsnd1,egsnd2,egsnd3,egsnd4} as well as measured the
branching ratios of the $\rho$, $\omega \to \etag$ decay in the 
$\eta \to 3\pi^0$ decay mode~\cite{egsnd4}.   
In this work we report on the measurement of the cross section 
of the process $\eeetag$ in the $\eta \to 3\pi^0$ decay mode in the 
broad c.m.energy range 600-1380~MeV as well as determination of the 
branching ratios of the $\rho$, $\omega$, $\phi \to \etag$ decay using 
the CMD-2 detector. 

\section{Experiment}

The general purpose detector \mbox{CMD-2} has been described in 
detail elsewhere~\cite{cmddec}. Its tracking system consists of a 
cylindrical drift chamber (DC) and double-layer multiwire proportional 
Z-chamber, both also used for the trigger, and both inside a thin 
(0.38~X$_0$) superconducting solenoid with a field of 1~T. 

The barrel CsI calorimeter with a thickness of 8.1~X$_0$ placed
outside  the solenoid has the energy resolution for photons of about
9\% in the energy range from 100 to 700~MeV. The angular resolution is 
of the order of 0.02 radians. The end-cap BGO calorimeter with a 
thickness of 13.4~X$_0$ placed inside the solenoid 
has the energy and angular resolution varying from 9\% to 4\% and from 
0.03 to 0.02 radians respectively for the photon energy in the range 
100 to 700~MeV.
The barrel and end-cap calorimeter systems cover a solid angle of
$0.92\times4\pi$ radians. 

The experiment was performed in the c.m. energy range 600-1380~MeV 
during the runs of 1997-2000. The analysis is based on the 
entire data sample corresponding to 26.3~pb$^{-1}$. 

The step of the c.m. energy scan varied from 500~keV near the $\omega$
and $\phi$ mesons peaks to 5~MeV far from the resonances.
The beam energy spread was about 400~keV at 1000~MeV.
Luminosity was measured using large angle Bhabha scattering events.

\section{Data analysis}

To study the process $\eeetag$, the decay mode $\eta \to 3\pi^0$ was 
chosen so that there are seven photons in the final state. 

Some of the photons can escape detection because of the threshold of 
shower detection, limited solid angle or shower merging.
An event can also have additional (``fake'') photons  because of the 
shower splitting, ``noisy'' crystals in the calorimeter and beam 
background.

At the first stage of analysis events were selected which had no tracks 
in the DC, from 6 to 8 photons, the total energy deposition
$E_{tot} > 1.5\, E_{beam}$, the total momentum 
$P_{tot} < 0.4\, E_{beam}$ and at least 3 photons detected in the
CsI calorimeter. The minimum photon energy was 20~MeV for the CsI and 
30~MeV for the BGO calorimeter.  
 
After kinematic reconstruction requiring energy-momentum  
conservation events with good 
reconstruction quality ($\chi^2 < 12$) and in which
for each photon the ratio of the reconstructed energy to the measured 
one was $0.7 < \omega_i\, / E_i < 1.9$ were retained.

The detection efficiency for the process under study was determined
by Monte Carlo simulation taking into account the neutral trigger 
efficiency~\cite{mast}. It rises steeply from 18\% at 600~MeV to 
about 33\% at the $\phi$ meson energy.
 
Since background conditions strongly vary in the energy range studied,
it was divided into three parts:
\begin{itemize}
\item
low energies --- $2E_{beam} <$ 950~MeV,
\item
the $\phi$ meson range --- 950~MeV $< 2E_{beam} <$ 1060~MeV,
\item
high energies --- $2E_{beam} >$ 1060~MeV.
\end{itemize}

Below the $\omega \pi^0$ production threshold ($2E_{beam} < 920$~MeV)
there are no other sources of multiphoton events with a significant 
cross section. Possible background from cosmic muons and QED processes 
is efficiently suppressed by the cut on the minimum number of photons 
and total energy deposition. All events meeting the selection criteria 
were considered as those of the process $\eeetag, \,\eta\to 3\pi^0$.

In the $\phi$ meson energy range the main background comes from the 
decay mode $\phi\to K^0_{L}K^0_{S}\to neutrals$. Since the cross section 
of this process is large, signal and background events are separated 
statistically. Outside the $\phi$ meson there is also a contribution
from the process $\ee\to\omega\pi^0$, $\omega\to\pi^0\gamma$.

Separation used the reconstructed recoil mass of the most energetic photon
(with the energy $\omega_1$) determined from
\beqn
M_{recoil}=\sqrt{(2E_{beam}-\omega_{1})^2-\omega_{1}^2}\, .
\eeqn       
 
\begin{figure}
  \includegraphics[width=\textwidth]{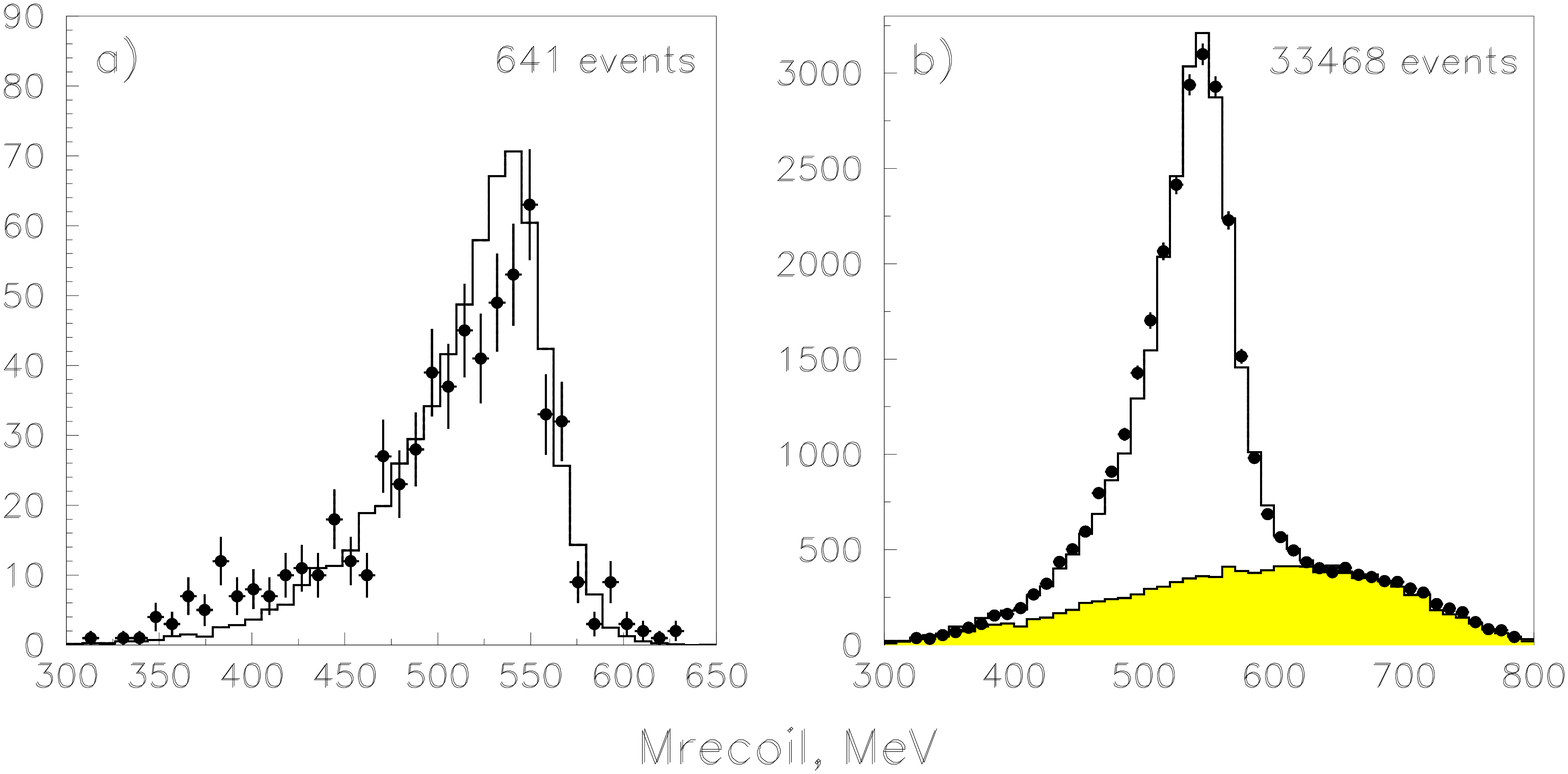}
  \vspace{-5mm}
  \caption{$M_{recoil}$ for the most energetic photon: 
 a) below 950~MeV, b) $\phi$ meson. 
        Dots correspond to experiment, solid histograms are simulation
of the $\etag$ production, the dashed histogram is background simulation.}
  \label{fig:recoil_mass}
\end{figure}

In Fig.~\ref{fig:recoil_mass} the distribution of the recoil mass for the 
most energetic photon at low energies and in the $\phi$ meson range is 
compared to the simulation. As expected, for $\etag$ events it 
exhibits a narrow peak at the $\eta$ meson mass of 547.3~MeV whereas
for background events a broad distribution is observed.

The shape of the background distribution was determined from 
events with bad reconstruction quality ($\chi^2 > 15$). To this end
the $M_{recoil}$ distribution was fit separately for events with six, 
seven and eight photons using the function
\beqn
F(z)=A_0 \cdot z^{\alpha} \cdot (1 - z)^{\beta}\, ,
\eeqn 
where $z=(x-x_{min})/(x_{max}-x_{min})$ and $x_{min}$ and $x_{max}$
are the spectrum boundaries.

For events satisfying the selection criteria, the same distribution
was fit by a sum of an asymmetric Gaussian and the background function
with the shape fixed as above. 
After that the shape 
of the functions was fixed and the fit was performed for all energies.   
 
At high energies where the background is much higher,
only events with seven photons were used.
A new kinematic fit requiring energy-momentum conservation was
performed with the following additional conditions.
Five photons with a smaller energy (soft photons) were selected 
and a kinematic fit to the $\pi^0\pi^0\gamma$ state was required. Thus, 
three free photons remained: the most energetic or the recoil one, 
a photon with the second energy and the one which
was free among the five soft. The invariant mass of the two 
latter photons $M_{\gamma\gamma}$ was used 
for further selection. 

After that, good reconstruction quality events ($\chi^2 < 7$) 
were selected in which for each photon 
$0.8<\omega_i\, /E_i<1.5$
and 
$50 < M_{\gamma\gamma} < 200$~MeV. 
With these cuts the detection efficiency above 1060~MeV is about 8.4\%. 

\begin{figure}
  \centering \includegraphics[width=.7\textwidth]{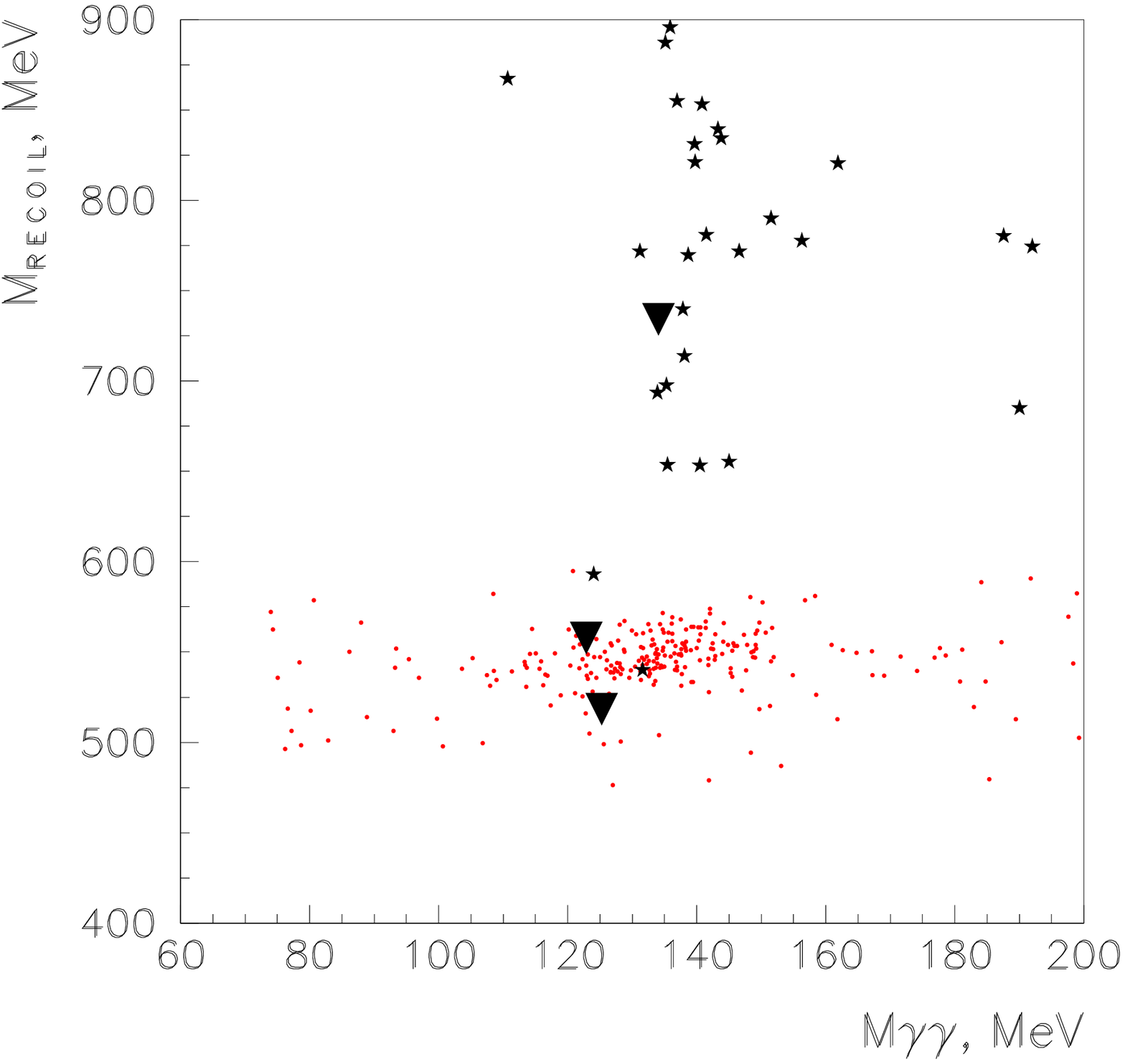}
  \vspace{-5mm}
  \caption{ $M_{recoil}$ for the most energetic photon versus 
    $M_{\gamma\gamma}$. Triangles correspond to experiment, dots are 
    the $\etag$ simulation, stars are the $\omega\pi^0\pi^0$ simulation.}
  \label{fig:e1_high}
\end{figure}

Three events have been selected at $2E_{beam} > 1300$~MeV. 
Figure~\ref{fig:e1_high}
shows the scatter plot of the recoil mass of the most energetic photon 
versus the invariant mass $M_{\gamma\gamma}$ of two free photons for
these events and simulation of the process $\eeetag$.

The main sources of background are the processes:
\beqn  \label{eq:ompi}
\ee \to \omega \pi^0 , \, \omega \to \pi^0 \gamma\, ,
\eeqn
and
\beqn \label{eq:ompipi}
\ee \to \omega \pi^0\pi^0 , \, \omega \to \pi^0 \gamma\, .
\eeqn    

The cross section of the process (\ref{eq:ompi})  is rather large 
(about 1.0-1.5~nb), however this process has five final photons only and 
is easily suppressed by the cut on the minimum number of photons.

The process (\ref{eq:ompipi})  has seven final photons, six of which 
pair to three $\pi^0$. Its events can be therefore separated 
only using the different dynamics of the $\etag$ and 
$\omega \pi^0 \pi^0$ final states. Its cross section is given by
\beqn
\sigma(\ee \to \omega\pi^0\pi^0 ,\, \omega \to \pi^0 \gamma) =
\frac{1}{2}\, \sigma(\ee \to \omega \pipi) \cdot 
B(\omega \to \pi^0 \gamma)\, ,
\eeqn      
so that in the energy range 1300-1400~MeV it is 0.01-0.02~nb according
to our measurement of the $\omega\pipi$ production~\cite{5pi_cmd}.

Other possible sources of the background are the radiative return
to the $\phi$ meson with a subsequent decay $\phi \to \eta \gamma$ and
the process $e^+e^- \to K^0_{L}K^0_{S}\pi^0$ which have a small cross 
section and can  be also suppressed  kinematically.

Simulation shows that in Fig.~\ref{fig:e1_high} events of the process  
$\eeetag$ should concentrate in the region $M_{recoil}=M_{\eta}$, 
$M_{\gamma\gamma}=M_{\pi^0}$ whereas those of the background have the 
broad distribution. Of three observed events two are in the signal region
and one is in the background region. 
The expected number of background events in the whole plot calculated 
from the cross section above and detection efficiency from Monte Carlo 
is equal to 1.3 consistent with the observation. This allows the
estimation of the contribution from the known background processes to the 
$\etag$ region at less than 0.1 event, so that the probability that
two observed events in the signal region come from background is less 
than 0.5\%.

\newpage

\section{Results}
\subsection{Approximation of the cross sections}

The energy dependence of the expected number of events is given by:
\begin{equation}
  \label{eq:Nth}
 N^{th}_{\etag}=L(\sqrt{s})\cdot \tilde{\sigma}(\sqrt{s}) \cdot
 \epsilon(\sqrt{s})\cdot B_{\eta\to 3\pi^0}\cdot B^3_{\pi^0\to\twog}\; ,
\end{equation}
where $L$ is the integrated luminosity at the c.m.energy 
$\sqrt{s}=2E_{beam}$, 
$\tilde{\sigma}$ is the visible cross section of the process $\eeetag$,
$\epsilon$ is the detection efficiency, 
$B_{\eta\to 3\pi^0}=(32.24 \pm 0.29)$\% and
$B_{\pi^0 \to 2\gamma}=(98.798 \pm 0.032)$\%
are the branching ratios of the corresponding decays~\cite{pdg}.  
The maximum likelihood method was applied to fit the experimental
data to the relation (\ref{eq:Nth}) with the parameterization of the
cross section described below. 
The fit took into account the beam energy spread 
and the uncertainty of the beam energy determination. 
The radiative corrections were calculated during the fit according 
to~\cite{radcor}: $\tilde{\sigma}(s)$ is the convolution of the Born 
cross section $\sigma(s)$ with a probability to emit photons,
often simply written as $\tilde{\sigma}(s)=\sigma(s) (1 + \delta)$
where $\delta$ is referred to as a radiative correction. 
The dependence of the detection efficiency on the energy 
of the emitted photon was determined from simulation.

The Born cross section of the process can be written as:
\begin{eqnarray}
  \label{eq:cross-section}
  \sigma_{\etag}(s)=\frac{F_{\etag}(s)}{s^{3/2}}\cdot
  \biggl|\sqrt{4\pi\alpha^2}C_\eta+
  \sum_{V=\rho, \omega, \phi, \rhop}A_V\biggr|^2\; , \\
  A_V=\sqrt{\sigma^{(0)}_{V}\frac{m_V^3}{F(m_V^2)}}\cdot
  \frac{m_V\Gamma_V e^{i\varphi_V}}{m^2_V-s-i\sqrt{s}\Gamma_V(s)} \, ,\nonumber
\end{eqnarray}
where $m_V$ is the mass of the resonance, $\Gamma_{V}(s)$ and 
$\Gamma_{V}=\Gamma_{V}(m_V^2)$ are its width at the squared c.m.energy s
and at the resonance peak respectively, $\delta_V$ is its relative phase,
$F(s)$ is a factor taking into account the energy
dependence of the phase space of the final state,  
$F_{\etag}(s) = p_{\eta}^3 =(\sqrt{s}(1 - m_\eta^2/2s))^3$, 
$C_{\eta}$ is a possible anomalous contribution, $\sigma_{V}^{(0)}$ is the
cross section at the resonance peak calculated without taking into 
account other contributions:
\begin{equation}
  \sigma_{V}^{(0)}=\sigma_{\ee\to V\to \etag}(m_V^2)=
  \frac{12\pi B_{V \to \ee} B_{V \to \etag}} {m^2_V}\; ,
\label{eq:sigma_G}
\end{equation}
where $B_{V\to \ee}$ and $B_{V\to\etag}$ are the corresponding
branching ratios.

\begin{table}
\caption{Energy, integrated luminosity, detection efficiency, number of 
selected events, radiative correction and Born cross section of the 
process $\ee\to\etag$.}
\begin{center}
\begin{tabular}{|c|c|c|c|c|c|}
\hline
$\sqrt{s}$, MeV& $L$, nb$^{-1}$& $\epsilon$& $N$& $\delta$& $\sigma$, nb 
\\\hline
  600 & 57.3 & 0.182 &   0 & -0.157 & $ < 0.84$ \\
  630 &  118 & 0.197 &   0 & -0.141 & $ < 0.37$ \\
  660 &  241 & 0.211 &   0 & -0.133 & $ < 0.17$ \\
  690 &  201 & 0.224 &   1 & -0.129 & $ 0.08^{+0.10}_{-0.06}$ \\
  720 &  430 & 0.237 &   9 & -0.125 & $ 0.32\pm 0.11$ \\
  750 &  216 & 0.249 &   7 & -0.116 & $ 0.47\pm 0.18$ \\
  760 &  211 & 0.253 &   5 & -0.114 & $ 0.34\pm 0.15$ \\
  764 & 40.7 & 0.255 &   1 & -0.115 & $ 0.35^{+0.46}_{-0.25}$ \\
  770 &  112 & 0.257 &   2 & -0.126 & $ 0.26^{+0.23}_{-0.14}$ \\
  774 &  200 & 0.259 &  14 & -0.149 & $ 1.02\pm 0.27$ \\
  778 &  204 & 0.260 &  22 & -0.189 & $ 1.64\pm 0.35$ \\
  780 &  199 & 0.261 &  23 & -0.204 & $ 1.79\pm 0.37$ \\
  781 &  262 & 0.261 &  31 & -0.204 & $ 1.84\pm 0.33$ \\
  782 &  646 & 0.262 & 103 & -0.197 & $ 2.45\pm 0.24$ \\
  783 &  282 & 0.262 &  48 & -0.181 & $ 2.56\pm 0.37$ \\
  784 &  345 & 0.262 &  54 & -0.158 & $ 2.28\pm 0.31$ \\
  785 &  204 & 0.263 &  24 & -0.133 & $ 1.66\pm 0.34$ \\
  786 &  195 & 0.263 &  33 & -0.107 & $ 2.31\pm 0.40$ \\
  790 &  153 & 0.264 &  10 & -0.032 & $ 0.82\pm 0.26$ \\
  794 &  183 & 0.266 &  14 &  0.000 & $ 0.92\pm 0.25$ \\
  800 &  268 & 0.268 &  13 &  0.014 & $ 0.57\pm 0.16$ \\
  810 &  253 & 0.272 &  10 &  0.017 & $ 0.46\pm 0.15$ \\
  820 &  303 & 0.275 &   8 &  0.017 & $ 0.30\pm 0.11$ \\
  840 &  618 & 0.282 &  26 &  0.014 & $ 0.47\pm 0.09$ \\
  880 &  385 & 0.294 &  13 & -0.001 & $ 0.37\pm 0.10$ \\
  920 &  470 & 0.305 &  10 & -0.035 & $ 0.23\pm 0.07$ \\
  940 &  336 & 0.310 &   8 & -0.052 & $ 0.26\pm 0.09$ \\
\hline
\end{tabular}
\end{center}
\label{tab:etag_cs1}
\end{table}

\begin{table}
\caption{Energy, integrated luminosity, detection efficiency, number of 
selected events, radiative correction and Born cross section of the 
process $\ee\to\etag$.}
\begin{center}
\begin{tabular}{|c|c|c|c|c|c|}
\hline
$\sqrt{s}$, MeV& $L$, nb$^{-1}$& $\epsilon$& $N$& $\delta$& $\sigma$, nb 
\\\hline
  950    &  232 & 0.313 &   3 & -0.064 & $ 0.14\pm 0.08$ \\
  958    &  257 & 0.315 &   7 & -0.075 & $ 0.30\pm 0.11$ \\
  970    &  256 & 0.318 &   5 & -0.095 & $ 0.23\pm 0.20$ \\
  984    &  403 & 0.321 &  18 & -0.125 & $ 0.50\pm 0.17$ \\
 1004    &  469 & 0.325 &  65 & -0.191 & $ 1.70\pm 0.28$ \\
 1010.3  &  486 & 0.326 & 127 & -0.226 & $ 3.32\pm 0.35$ \\
 1015.7  &  442 & 0.327 & 534 & -0.269 & $15.97\pm 0.88$ \\
 1016.78 & 1040 & 0.328 &1998 & -0.278 & $25.64\pm 0.80$ \\
 1017.82 & 1588 & 0.328 &4278 & -0.282 & $36.71\pm 0.90$ \\
 1018.68 & 1540 & 0.328 &5773 & -0.275 & $52.14\pm 0.93$ \\
 1019.69 & 1456 & 0.328 &5509 & -0.240 & $50.10\pm 0.96$ \\
 1020.58 &  924 & 0.328 &2826 & -0.183 & $36.45\pm 1.02$ \\
 1021.51 &  477 & 0.328 & 955 & -0.106 & $21.49\pm 0.97$ \\
 1022.74 &  405 & 0.329 & 516 &  0.009 & $12.09\pm 0.72$ \\
 1027.8  &  530 & 0.330 & 212 &  0.577 & $ 2.45\pm 0.39$ \\
 1033.7  &  520 & 0.331 &  98 &  1.424 & $ 0.75\pm 0.28$ \\
 1039.6  &  582 & 0.332 &  66 &  2.503 & $ 0.31\pm 0.22$ \\
 1049.7  &  452 & 0.334 &  35 &  5.075 & $ 0.12^{+0.21}_{-0.12}$ \\
 1060    &  569 & 0.335 &  24 &  8.552 & $ 0.04^{+0.17}_{-0.04}$ \\
 1100    &  893 & 0.084 &   3 &  16.18 & $ 0.01^{+0.08}_{-0.01}$\\
 1160    & 1250 & 0.084 &   0 &  4.530 & $ < 0.07$ \\
 1224    &  916 & 0.084 &   0 &  0.136 & $ < 0.08$ \\
 1290    & 1672 & 0.084 &   0 & -0.130 & $ < 0.06$ \\
 1354    & 1777 & 0.084 &   2 & -0.149 & $ 0.05^{+0.04}_{-0.03}$ \\
\hline
\end{tabular}
\end{center}
\label{tab:etag_cs2}
\end{table}

The Gounaris-Sakurai model has been used for the description of the 
$\rho$ meson~\cite{gounaris}.
To describe the energy dependence of the  $\omega$ and $\phi$ meson 
width their main decay modes  $\pipipi$, $\pig$ as well as
$K^0_{L}K^0_{S}$, $K^+K^-$, $\pipipi$ and $\etag$
respectively were taken into account as in~\cite{kuzmin}.
For the $\rhop$ the energy dependence of the width assumed 60\% and
40\%  branching ratios for its decays into $a_1(1260)\pi$ and
$\omega\pi$ respectively~\cite{root}.

The detailed information about the experiment including the cross section 
of the process $\eeetag$ and the radiative
correction obtained from the fit is presented  in 
Tables~\ref{tab:etag_cs1},~\ref{tab:etag_cs2}. The last five energy 
points combine the whole data sample at $2E_{beam}>1080$~MeV analyzed 
with stricter selection criteria. No events were selected in the
energy ranges 600 to 660~MeV and 1160 to 1290~MeV, therefore our
results are presented as upper limits at 90\% C.L. Events selected in 
the energy range 1025-1100~MeV are mostly due to the radiative return
to the $\phi$ meson as clear from a very high value of the radiative 
correction $\delta$, so that the uncertainties of the cross section
are very large.   

\subsection{Results of the fits} 

In  all the following fits the cross sections at the resonance peaks 
$\sigma_{\rho}^{(0)}$, $\sigma_{\omega}^{(0)}$, $\sigma_{\phi}^{(0)}$ 
as well as the  $\phi$ meson mass 
$m_{\phi}$  are free parameters. Unless otherwise stated,   
the $\rho$ meson phase is chosen to be  $\varphi_\rho=0^\circ$ while
those for the $\omega$ and $\phi$ mesons are 
$\varphi_\omega=\varphi_\rho$, $\varphi_\phi=\varphi_\rho+180^\circ$ 
in agreement with the quark model. 
The values of other parameters
are taken from~\cite{pdg}, those of the $\rho(770)$ meson from our 
measurement~\cite{rho}. 

Two different models were considered: the first one is VDM with the
$\rho$, $\omega$, $\phi$ mesons and in the second one an additional 
$\rhop$ meson was included. In both cases $C_{\eta}$=0.

\begin{figure}
  \includegraphics[width=\textwidth]{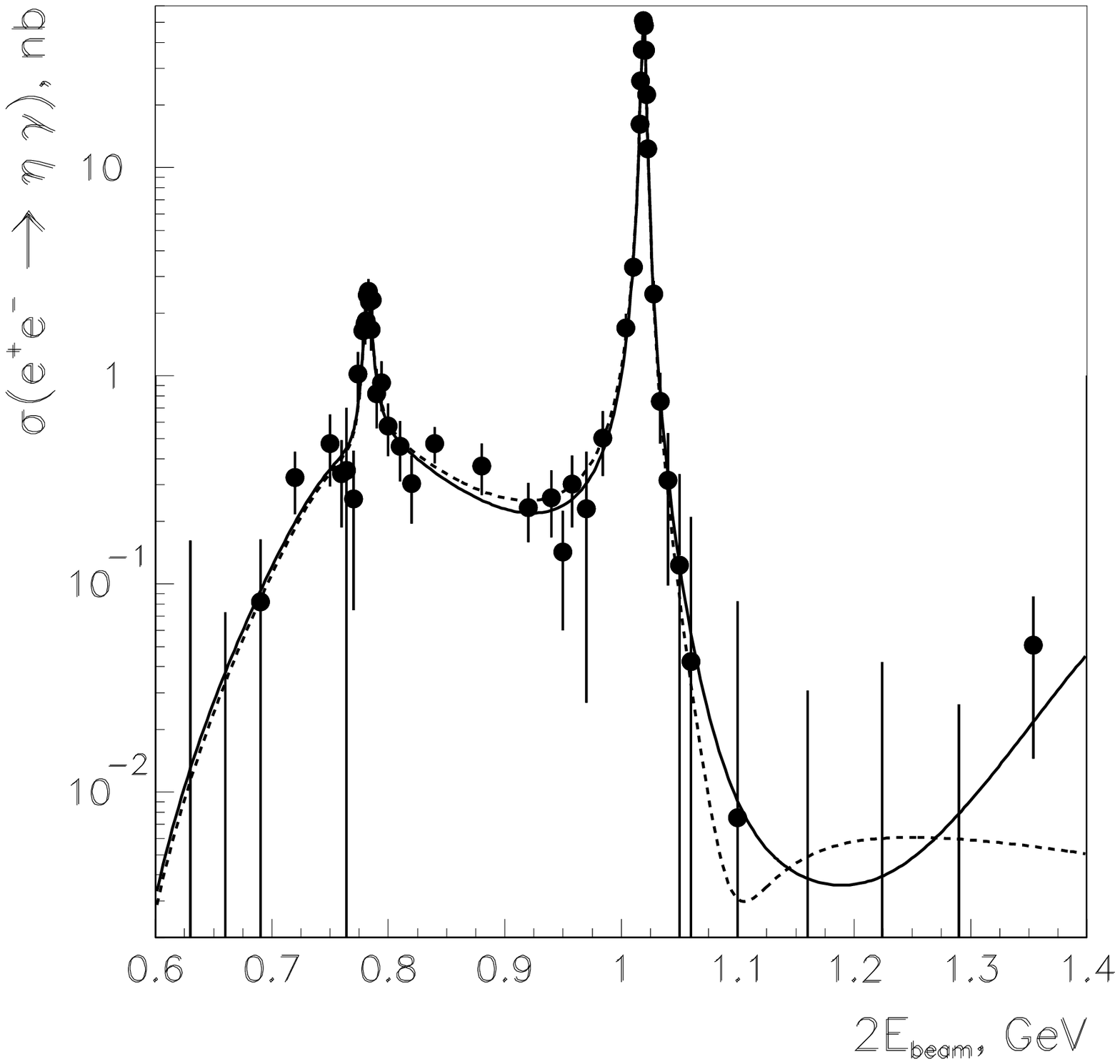}
  \vspace{-9mm}
  \caption{$\eeetag$ cross section in the optimal fits: 
    the dashed curve is VDM, the solid one is VDM + $\rhop\to\etag$.}
  \label{fig:cross-section}
\end{figure}

Results of the fits are shown in Table~\ref{tab:BB} in terms of 
the product of the branching ratios $B_{V \to \ee}B_{V \to \etag}$
determined from $\sigma_{V}^{0}$ according to (\ref{eq:sigma_G}) as
well as in Fig.~\ref{fig:cross-section} together with the fit curves.

\begin{table}
\caption{ $B_{V \to \ee} B_{V \to \etag}$ in various models.}
\begin{center}
\begin{tabular} {|c|c|c|}
\hline
                        & VDM             & VDM + $\rhop$  \\
\hline
$\rho, 10^{-8}$         & $1.51 \pm 0.18 \pm 0.10$ & 
                          $1.61 \pm 0.20 \pm 0.11$  \\
\hline
$\omega, 10^{-8}$       & $3.62 \pm 0.52 \pm 0.22$ & 
                          $3.41 \pm 0.52 \pm 0.21$  \\
\hline
$\phi, 10^{-6}$         & $3.837 \pm 0.041 \pm 0.158$ & 
                          $3.850 \pm 0.041 \pm 0.159$ \\
\hline
$\rhop, 10^{-9}$        & $\equiv 0$ & $3.7^{+2.9}_{-2.0}$ \\
\hline
$m_{\phi}$, MeV             & $1019.42 \pm 0.04 \pm 0.05$ &  
                              $1019.40 \pm 0.04 \pm 0.05$ \\
\hline
$\chi^2/$n.d.f.         & 40.8 / 47       & 36.4 / 46  \\
\hline
\end{tabular}
\label{tab:BB}
\end{center}
\end{table}

Although the fit quality ($\chi^2$/n.d.f.) in VDM is good, from 
Fig.~\ref{fig:cross-section}
it is clear that at $2E_{beam}>1300$~MeV the measured cross section 
is not described by the contributions of  $\rho$, $\omega$ and 
$\phi$ mesons only. This fact provides evidence for possible
additional contributions coming for example from the higher resonances.

In the second model the $\rhop$
mass and width were fixed at 1450~MeV and 310~MeV respectively
and the relative phase of the $\rhop$ was fixed at the value 
of $\varphi_{\rho}$.     
From Table~\ref{tab:BB} it can be seen that the agreement with the data 
is slightly better, mostly due to the high energy point. 

One more possibility is to take into account a possible anomalous
contribution by adding a nonresonant term determined 
by the amplitude $C_\eta$ in the formula (\ref{eq:cross-section}).
$C_{\eta}$ is related to the width of the $\eta\to\twog$ 
decay~\cite{ben}. Although this model also provides a good description
of the data in the whole energy range including the highest energy point,
the high energy behaviour of the anomalous contribution is unknown 
and the model in its current form is hardly applicable above 1 GeV. 

In all models considered the optimal values of the $m_{\phi}$
are consistent within errors with the world average value, thus 
confirming the correctness of the energy scale.
  
\subsection{Systematic errors}

The main sources of  systematic uncertainties in the cross section 
determination are listed below: 

\begin{itemize}
\item
Stability to the variation of selection criteria --- 5\% below 950~MeV,
3.5\% from 950~MeV to 1060~MeV, 8\% above 1060~MeV;
\item
The choice of the function describing the recoil photon spectrum at the
$\phi$ meson --- 1\%;
\item
The error of the detection efficiency including the neutral trigger 
efficiency --- 1\%;
\item
Determination of the integrated luminosity --- 1\% in the $\phi$ meson
energy range, 2\% at the high energies and 3\% below 950~MeV.
\item
Radiative corrections --- 1\% for the $\omega$, $\phi$ and 3\% for the 
$\rho$ meson.
\end{itemize}

For the determination of the cross section one should additionally 
take into account the uncertainty of $B_{\eta \to 3\pi^0}$ equal to 
0.9\%.
Then the systematic uncertainty of the cross sections quoted in 
Tables~\ref{tab:etag_cs1},~\ref{tab:etag_cs2} is 6.1\% below 950~MeV,
4.1\% at the $\phi$ meson and 8.4\% above 1060~MeV. 
      
For the determination of the  branching fractions $B(V \to \etag)$
one should additionally take into account the uncertainty of 
$B_{V \to \ee}$ which is 1.8\% for the $\rho$~\cite{rho} while it is 
2.7\% for the $\omega$ and $\phi$~\cite{pdg}.
As a result, the systematic uncertainty is 6.9\%, 6.6\% and 4.9\%
for $\rho$, $\omega$ and $\phi$.

\section{Discussion}

Although the product of the branching ratios for the $\rhop$ is small
and differs from zero by 1.5 standard deviations only, the observed 
events at high energy
can hardly be explained by the known background processes. Moreover, 
one can independently estimate the cross section of the process
$\eeetag$ from the measured cross section of the process
$\eeetapipi$~\cite{5pi_cmd}. The latter process is
known to proceed via the $\eta\rho$ state and one can estimate the 
$\rho \to \gamma$ transition from VDM so that  
\begin{equation}
  \label{etarho_etagamma}
  \sigma_{\eta\gamma}(E) = \sigma_{\eta\rho}(E)\cdot
 \frac{4\pi\alpha}{f_\rho^2}\cdot\frac{F_{\eta\gamma}(E)}{F_{\eta\rho}(E)} \, ,
\end{equation}
where 
$f_\rho$ is a coupling constant of the $\rho$ meson with a photon which
can be obtained from the decay width
$\Gamma_{\rho ee}=4\pi m_\rho \alpha^2/3f^2_\rho$;
$F_{\eta X}(E)$ is a phase space factor for the $\eta X$ final state.
The value $\sigma_{\eta\gamma}(1350~\mbox{MeV})=0.022\pm0.004$~nb 
obtained from (\ref{etarho_etagamma}) at $\varphi_\rhop=\varphi_\rho$ 
is not incompatible with the measured cross section 
$\sigma_{\etag}(1350~\mbox{MeV})=0.051^{+0.044}_{-0.028}$~nb. 

Therefore we consider our observation as evidence for the existence 
of the $\rhop \to \eta \gamma$ decay. Since our measurement covers
the energy range below 1380~MeV, i.e. the left slope of the 
$\rhop$ meson only, we can not determine its parameters. However,
one can obtain the $\eeetag$ cross section in  the much broader energy
range 1300 to 2450~MeV from the measurements of the cross section 
$\eeetapipi$  by DM2~\cite{5pi_dm2} and CMD-2~\cite{5pi_cmd} using 
(\ref{etarho_etagamma}). If the cross sections obtained in such
a way are combined with the cross section $\eeetag$ from our measurement
and a fit is performed with a free mass and width of the  $\rhop$ meson, 
then  reasonable description of the whole data is achieved with the 
following values of resonance parameters:

\begin{eqnarray}
  \label{eq:BB_rhop}
  B_{\rho\to\ee}\cdot B_{\rho\to\etag} & = & 
  (1.52 \pm 0.17 \pm 0.10)\cdot 10^{-8}\, ,\\
  B_{\omega\to\ee}\cdot B_{\omega\to\etag} & = &
  (3.60 \pm 0.51 \pm 0.22)\cdot 10^{-8}\, ,\nonumber \\
  B_{\phi\to\ee}\cdot B_{\phi\to\etag} & = &
  (3.849 \pm 0.040 \pm 0.159)\cdot 10^{-6}\, ,\nonumber \\
  B_{\rhop\to\ee}\cdot B_{\rhop\to\etag} & = &
  (10.0\pm 2.2)\cdot 10^{-9}\, ,\nonumber \\
  m_{\phi} & =&  1019.40 \pm 0.04 \pm 0.05\,\mbox{MeV}\, ,\nonumber \\
  m_{\rhop} & = & 1497 \pm 14\,\mbox{MeV}\, ,\nonumber \\
  \Gamma_{\rhop} & = & 226\pm 44\,\mbox{MeV}\, ,\nonumber \\
  \chi^2/\,\mbox{n.d.f.} & = & 68.9\, /\, 71\, .\nonumber
\end{eqnarray}

\begin{figure}
  \centering \includegraphics[width=.9\textwidth]{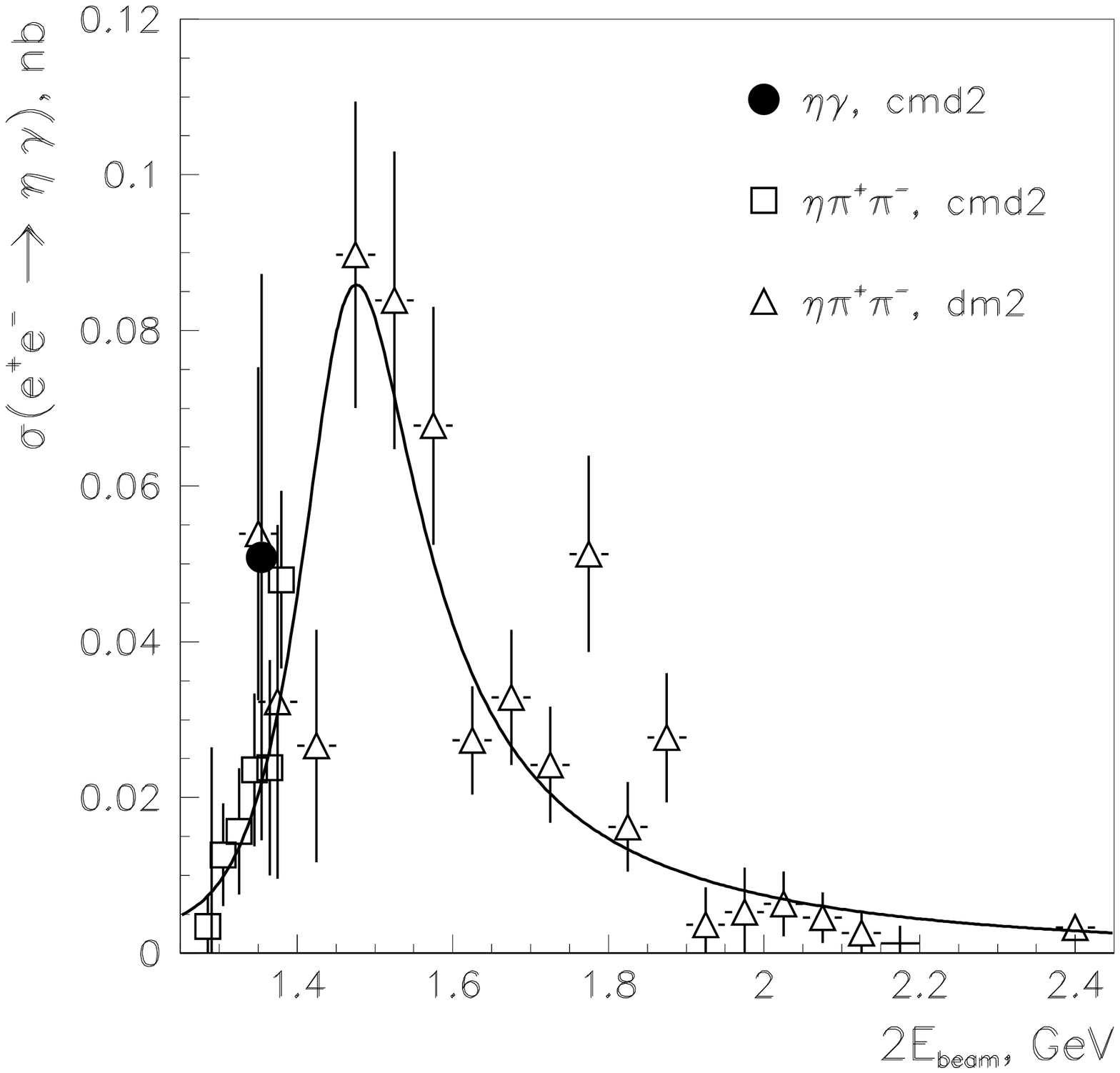}
  \vspace{-6mm}
  \caption{Cross sections: dots are $\eeetag$, squares and triangles
   are results of the  calculation from the $\eeetapipi$ production
   for CMD-2 and DM2 respectively.}
  \label{fig:rhop_cs}
\end{figure}

The cross section $\eeetag$ and the obtained fit curve  are also shown 
in Fig.~\ref{fig:rhop_cs} for the energies above 1200~MeV.

Analysis of the fit results in Table~\ref{tab:BB}  as well as those 
in (\ref{eq:BB_rhop}) shows that the model dependence of the
branching ratios does not exceed 5\% for the $\rho$ and $\omega$
mesons whereas for the $\phi\to\etag$ decay it is less than  1\%. 
The fit with the $\rhop$ meson gives better description of the data
in the whole energy range studied. In all fits described the value 
of $\chi^2/\,\mbox{n.d.f.}$ calculated using our measurement only is
practically the same. 

The $\rhop$ existence has been well established before in various 
decay modes~\cite{pdg} and as follows from the analysis above, it can 
also decay into $\etag$.  
Therefore we present the final results in the model with the  $\rhop$
meson using both our measurement and the calculation from the
cross section of the process $\eeetapipi$. 

The values of the cross section obtained allow the conclusion 
that above the $\phi$ meson 
the contribution of the $\etag$ production to the total hadronic 
cross section is negligibly small.  

One can use the leptonic branching ratios measured
independently to determine the absolute branching ratio 
$B_{V \to \etag}$. 

Table~\ref{tab:etag_rhoom} compares the branching ratios of the 
$\rho$, $\omega \to\etag$
decays obtained in this work with the results of SND~\cite{egsnd4} 
as well as with the world average values~\cite{pdg}.

\begin{table}
\caption{Branching ratios for $\rho$, $\omega \to \etag$ decays.}
\begin{center}
\begin{tabular}{|l|c|c|}      \hline
Experiment & $B(\rho\to\etag), 10^{-4} $ & $B(\omega\to\etag), 10^{-4}$ \\ 
\hline
  PDG, 2000 \cite{pdg} &  $2.4^{+0.9}_{-0.8}$ & $6.5\pm 1.0$ \\
 \hline
SND, 2000 \cite{egsnd4} & $2.73\pm 0.31\pm 0.15$& $4.62\pm 0.71\pm 0.18$\\
 \hline
This work, 2001 & $3.28 \pm 0.37 \pm 0.23$ & $5.10 \pm 0.72 \pm 0.34$  \\
 \hline

\end{tabular}
\label{tab:etag_rhoom}
\end{center}
\end{table}

Table~\ref{tab:etag_phi} compares the branching ratio of the 
$\phi\to\etag$
decay obtained in this work with the results of the measurements
performed before 1998 and listed in the 1998 edition of the Review 
of Particle Physics~\cite{pdg98} as well as with the recent 
results obtained by two groups at VEPP-2M. 

\begin{table}
\caption{Branching ratio of $\phi \to \etag$ decay.}
\begin{center}
\begin{tabular}{|l|c|c|}      \hline
Experiment &  $\eta$ decay channel & $B(\phi\to\etag), \% $ \\ \hline
   PDG, 1998 \cite{pdg98} & Average & $1.26\pm 0.06$ \\ \hline
  SND, 1998 \cite{egsnd1} & $\eta\to3\pi^0$     &  
  $1.246\pm 0.025\pm 0.057$ \\ 
  CMD-2, 1999 \cite{egcmd} & $\eta\to\pipipi$ & 
  $1.18\pm 0.03\pm 0.06$ \\  
  SND, 1999 \cite{egsnd2} & $\eta\to\twog$ &  
  $1.338\pm 0.012\pm 0.052$ \\ 
  SND, 2000 \cite{egsnd3} & $\eta\to\pipipi$ &  
  $1.259\pm 0.030\pm 0.059$\\ 
  SND, 2000 \cite{egsnd4} & $\eta\to 3\pi^0$ &  
  $1.353\pm 0.011\pm 0.052$\\ \hline
  This work, 2001 & $\eta\to 3\pi^0$ & $1.287\pm 0.013\pm 0.063$ \\ \hline
\end{tabular}
\label{tab:etag_phi}
\end{center}
\end{table}

The results of high precision studies of the $\phi \to \etag$ decay
can be also used to improve the precision of our knowledge of the relative
branching ratios of the $\eta$ meson. It is known that data on 
the main decay modes of the $\eta$ meson from different experiments
($\eta\to\twog$, $\eta\to3\pi^0$, $\eta\to\pipipi$)~\cite{pdg} 
are not consistent with each other and their averaging requires
a scale factor of 1.2-1.3. CMD-2 performed two independent
measurements of the $B(\phi \to \etag)$ using $\eta$ decays into
$\pipipi$~\cite{egcmd} and $3\pi^0$ allowing the determination 
of the ratio $B(\eta\to 3\pi^0)/B(\eta\to\pipipi)$.
In such a ratio some of the systematic uncertainties cancel
(e.g. the error of $B_{\phi\to\ee}$, uncertainty of luminosity 
determination).

Using the value of 
$B(\phi\to\etag)=(1.18\pm 0.03\pm 0.06)~\%$, 
obtained in~\cite{egcmd}, one can determine
\begin{equation}
  \label{eq:comp_cmd_3pi}
  \frac{B(\eta\to 3\pi^0)}{B(\eta\to\pipipi)}=1.52\pm 0.04\pm 0.08\, ,
\end{equation}
which is compatible with all  previous measurements~\cite{pdg} and 
has better accuracy. The accuracy of our result
is also better than that of the world average:  
\begin{equation}
\label{eq:comp_pdg_3pi}
 \frac{B(\eta\to 3\pi^0)}{B(\eta\to\pipipi)}=1.34\pm 0.10\, . 
\end{equation}

\section{Conclusions}

The following results were obtained by the CMD-2 detector:
\begin{itemize}
\item
Using a data sample corresponding to the integrated luminosity of 
26.3~pb$^{-1}$ the cross section of the process $\ee \to \etag$ was
measured in the c.m.energy range 600-1380~MeV.
\item
The Vector Dominance Model extended by introducing a $\rhop$ meson
decaying to $\etag$ can well describe the data. 
The cross sections in the peak as well as the
branching ratios of $\rho$, $\omega$ and $\phi$ decay into
$\etag$ were determined. Evidence for the $\rhop \to \eta \gamma$ 
decay was obtained for the first time. 
\item
Measurement of the $B(\phi \to \etag)$ by one detector in two
decay modes of the $\eta$ allows the determination of the ratio:
$$
\frac{B(\eta\to 3\pi^0)}{B(\eta\to\pipipi)}=1.52\pm 0.04\pm 0.08\, ,
$$
compatible with the world average and surpassing it in accuracy.
\end{itemize}

The authors are grateful to the staff of VEPP-2M for the
excellent performance of the collider, to all engineers and 
technicians who participated in the design, commissioning and operation
of CMD-2. We acknowledge useful and stimulating discussions with 
N.N.Achasov, M.Benayoun, A.V.Berdyugin and V.L.Chernyak.

\end{document}